\documentclass[preprint,prc,aps,nofootinbib,superscriptaddress]{revtex4}
\usepackage{graphicx}
\usepackage{epsfig}
\usepackage{bm}
\usepackage{amssymb}

\newcommand{\be}{\begin{eqnarray}}
\newcommand{\ee}{\end{eqnarray}}
\begin{document}

\title{Magnetic moments of octet baryons at finite density and temperature}

\author{C. Y. Ryu} 
\affiliation{Department of Physics,
Soongsil University, Seoul 156-743, Korea}

\author{C. H. Hyun} 
\affiliation{Department of Physics Education, Daegu University,
Gyeongsan 712-714, Korea}

\author{Myung-Ki Cheoun} \email{cheoun@ssu.ac.kr}
\affiliation{Department of Physics,
Soongsil University, Seoul 156-743, Korea}


\begin{abstract}
We investigate the change of magnetic moments of octet baryons in
nuclear matter at a finite density and temperature. Quark-meson
coupling models are employed in describing properties of octet
baryons and their interactions. Magnetic moments of octet baryons
are found to increase non-negligibly as density and temperature
increase, and we find that temperature dependence can be strongly
correlated with the quark-hadron phase transition. Model
dependence is also examined by comparing the results from the
quark-meson coupling (QMC) model to those by the modified QMC
(MQMC) model where the bag constant is assumed to depend on
density. Both models predict sizable dependence on density and
temperature, but the MQMC model shows a more drastic change of
magnetic moments. Feasible changes of the nucleon mass by
strong magnetic fields are also reported in the given models.

\end{abstract}
\maketitle

\section{Introduction}
Recently, the magnetic moment of a $\Lambda$ in $^7_\Lambda$Li was
observed in BNL to test its medium modification. It shows us that
the magnetic moment may be changed in nuclear medium although the
error is still very large \cite{tamura07}. The experiment might be
able to extend to the heavy ion collision which can make hot and
dense nuclear matter. In this context, it would be interesting to
study change of magnetic moments of baryons in hot and dense
matter. The investigation for the change of baryon properties in
hot and dense matter is important in the interpretation of many
exotic phenomena occurring in the proto-neutron star and the heavy
ion collision. The subject has been studied by various models such
as relativistic mean fields (RMF) models \cite{zakout05,kla06},
chiral motivated models \cite{yazaki89,dex08}, and so on. In
specific, the quark-meson coupling (QMC) model \cite{guichon88},
one of the RMF models, is found to effectively describe exotic
nuclear matter as well as finite nuclei.

The QMC model is one of the extension of quantum hadrodynamics
(QHD) in which interactions between baryons are mediated by the
exchange of $\sigma$ and $\omega$ mesons, describing the
attraction and the repulsion, respectively. But, in the QMC model,
quarks inside baryons interact directly with meson fields. One of
merits is that one can evaluate properties of baryons with quark
degrees of freedom. For instance, effective masses of baryons in
nuclear medium can be obtained from the calculation of the MIT bag
by considering quark energies. Likewise, magnetic moments of
baryons can be also calculated with SU(6) quark wave functions and
a bag radius as well. In the present work, we investigate effective
masses and magnetic moments of baryon octet in hot and dense
matter.

Under the assumption that the matter reaches to thermal
equilibrium, the matter can be described through the minimum of a
thermal grand potential or the maximum of the pressure. Applying
the maximum condition of the pressure at finite temperature,
$\sigma$ and $\omega$ mesons can be self-consistently determined,
giving rise to the change of magnetic moments of baryons. 
Our results may be relevant to the QCD phase transition and
medium effect in high-energy nuclear collisions.

The paper is organized as follows. In Sec. II, the QMC model for
hot and dense matter is briefly explained. Magnetic moments of
baryons are obtained from SU(6) quark wave functions. Results and
discussions are followed in Sec. III. Sec. IV is devoted to the 
summary.

\section{Model}
Since one can find details of the description of hot and dense
nuclear matter with the QMC model in Ref.~\cite{panda03}, in this
section, we address the most essential ingredients of the model.
If we ignore the excitation of quarks in a baryon, the QMC model
treat a nucleon as a MIT bag in hot and dense nuclear matter. The
quark field $\psi_q$ inside the bag satisfies the Dirac equation
\be \left[ i \gamma \cdot \partial - ( m_q - g^q_\sigma\, \sigma)
- g^q_\omega\, \gamma^0 \, \omega_0 \right]\, \psi_q = 0,
\ee
where $m_q$ $\left( q= u,\, d,\, s\right)$ is the bare quark mass,
$\sigma$ and $\omega_0$ are the mean fields of the $\sigma$ and
$\omega$ mesons, respectively, and $g^q_\sigma$ and $g^q_\omega$
are coupling constants between quarks fields and the meson fields. 
We assume $m_u=m_d=0$ and $m_s = 150$ MeV for bare quark masses.

The ground state solution of the Dirac equation is given by
\be
\psi_q(\bm{r},\, t) =
{\cal N}_q \exp(-i \epsilon_q t/ R)
\left(
\begin{array}{c}
j_0(x_q\, r/R) \\
i\, \beta_q\, \bm{\sigma}\cdot\hat{\bm{r}}\, j_1(x_q\, r/R)
\end{array}
\right)
\frac{\chi_q}{\sqrt{4 \pi}},
\label{eq:quarkwf}
\ee
with
\be
{\cal N}^{-2}_q&=&2 R^3 j_0^2(x_q)[\Omega_q (\Omega_q -1)+R\,m_q^* /2]/x^2_q,\\
\epsilon_q &=& \Omega_q + g^q_\omega\,\omega_0 \, R, \\
\beta_q &=& \sqrt{\frac{\Omega_q - R\, m^*_q}{\Omega_q + R\, m^*_q}}, \\
\Omega_q &=& \sqrt{x^2_q + (R\, m^*_q)^2}, \\
m^*_q &=& m_q - g^q_\sigma\, \sigma, \ee
where $R$ is the bag radius. $j_0(x)$ and $j_1(x)$ are the
spherical Bessel functions, and $\chi_q$ is the quark spinor. The
value of $x_q$ is determined from the boundary condition on the
bag surface
\be j_0(x_q) = \beta_q\, j_1(x_q).
\label{eq:boundcond} \ee
The energy of a baryon with ground state quarks is given by
\be E_b &=& \sum_q \frac{\Omega_q}{R_b} - \frac{Z_b}{R_b} +
\frac{4\pi}{3} R_b^3 B_b, \label{eq:bagery}
\ee
where $B_b$ is the bag constant, and $Z_b$ is a phenomenological
constant introduced to take into account the zero-point motion of
the baryon. Subscript `$b$' denotes species of a baryon.
In the QMC model, the bag constant $B_b$ is independent of density
and temperature, whereas in the modified QMC (MQMC) model it is
assumed to depend on density and temperature. In this work, we
employ the direct coupling form given in Ref.~\cite{jin96},
\be
B_b(\sigma) = B_{b0} {\rm exp} \left(-\frac{4g_\sigma^b
\sigma}{m_N}\right).
\ee
As shown later on, $\sigma$-field is a function of density and
temperature, so that the bag constant $B_b$ depends on density and
temperature in the MQMC model. Effective mass of a baryon $b$ in
hot and dense matter is given by
\be m^*_b = \sqrt{E^2_b - \sum_q \left(\frac{x_q}{R_b} \right)^2}.
\label{eq:efmass} \ee
The value of bag radius in free space is usually chosen
close to the radius of the nucleon charge form factor.
In this work, we choose $R_b=1.0$~fm for free-space bag radius.
We determine the remaining parameters $B_{b0}$ and $Z_b$ to
reproduce the empirical value of the baryon mass in free space with 
the minimization condition
\be \frac{\partial m_b^*}{\partial R_b} = 0. \label{eq:minimal}
\ee
Numerical values of $B_{b0}$ and $Z_b$ for each baryon can be found 
in Ref.~\cite{ryu09}.
Quark-meson coupling constants
$g^q_\sigma$, $g^q_\omega$ and $g^b_\sigma$ are fitted to
reproduce the binding energy per a baryon in infinite nuclear
matter (16 MeV) and reasonable compression modulus ($\sim$ 280
MeV) at the saturation density ($\rho_0 = 0.17\, {\rm fm}^{-3}$)
under zero temperature. Numerical values of the coupling constants
are given in Ref.~\cite{RHHK}.

In hot matter, pairs of nucleon and antinucleon are produced and
thus the energy density is calculated as
\be
\varepsilon=\sum_b \frac{\gamma}{(2\pi)^3} \int d^3 k \sqrt{k^2 +
{m_b^*}^2} (f_b + \bar f_b) + \frac 12 m_\omega^2 \omega_0^2 +
\frac 12 m_\sigma^2 \sigma^2,
\ee
where $\gamma$ is the spin-isospin degeneracy factor. Functions
$f_b$ and $\bar f_b$ are the Fermi-Dirac distributions for baryons
and antibaryons
\be
f_b &=& \frac{1}{e^{(\epsilon_b^* - \mu_b^*)/T} + 1}, \\
\bar f_b &=& \frac{1}{e^{(\epsilon_b^* + \mu_b^*)/T} + 1},
\ee
where $\epsilon_b^* = \sqrt{k^2 + {m_b^*}^2}$ is the effective
energy and $\mu_b^* = \mu_b - g_{\omega b} \omega_0$ is the
effective chemical potential. If a baryon density $\rho_b$ is
given, we can determine the chemical potential $\mu_b$ from
\be
\rho_b = \frac{\gamma}{(2\pi)^3} \int d^3 k ( f_b - \bar f_b ),
\label{eq:baryon_density}
\ee
where $\omega_0$-meson field in $f_b$ and $\bar f_b$ is determined
by
\be \omega_0 = \sum_b \frac{g_{\omega b}}{m_\omega^2} \rho_b.
\label{eq:omega}
\ee
Pressure, which is the negative of the grand thermodynamic
potential density, is given by
\be P  = \sum_b
\frac 13 \frac{\gamma}{(2\pi)^3} \int d^3 k \frac {k^2}{\sqrt{k^2
+ {m_b^*}^2}} (f_b + \bar f_b) + \frac 12 m_\omega^2 \omega_0^2 -
\frac 12 m_\sigma^2 \sigma^2.
\ee
Mean-field value for the scalar meson $\sigma$ is determined
through the minimization of the thermodynamic potential, or
equivalently maximizing the pressure with respect to the field.
Maximization of $P(m_b^*, \sigma)$ with respect to $\sigma$ fields
can be written as \be \frac{d P}{d \sigma} = \sum_b \frac{\partial
m_b^*}{\partial \sigma} \left(\frac{\partial P}{\partial
m_b^*}\right)_{\mu_b, T} + \left ( \frac{\partial P}{\partial
\sigma}  \right )_{m_b^*} = 0, \label{eq:maximum} \ee where \be
\left( \frac{\partial P}{\partial \sigma}  \right)_{m_b^*} = -
m_\sigma^2 \sigma, \ee and \be \left( \frac{\partial P}{\partial
m_b^*} \right )_{\mu_b, T} &=& - \frac{\gamma}{3}
\frac{1}{(2\pi)^3} \int d^3 k \frac{k^2}{{\epsilon_b^*}^2}
      \frac{m_b^*}{\epsilon_b^*}[f_b + \bar f_b] \nonumber \\
&& -\frac{\gamma}{3} \frac{1}{(2\pi)^3}\frac{1}{T} \int d^3 k
    \frac{k^2}{\epsilon_b^*} \frac{m_b^*}{\epsilon_b^*}
[f_b ( 1-f_b ) + \bar f_b (1-\bar f_b)] \nonumber \\
&& -\frac{\gamma}{3} \frac{1}{(2\pi)^3} \frac{1}{T} g_{\omega b}
    \left( \frac{\partial \omega_0}{\partial m_b^*} \right )_{\mu_b, T}
    \int d^3k \frac{k^2}{\epsilon_b^*} [f_b (1-f_b) - \bar f_b (1-\bar f_b)]
    \nonumber \\
&& + m_\omega^2 \omega_0
\left(\frac{\partial\omega_0}{\partial m_b^*} \right)_{\mu_b, T}.
\label{eq:scalar_density}
\ee
Since the baryon density $\rho_b$ and temperature $T$ are treated
as input parameters, the variation of the vector mean field $\omega_0$ with
respect to the effective baryon mass $m_b^*$ at a given value of the baryon
density $\rho_b$ in Eqs.~(\ref{eq:baryon_density}) and (\ref{eq:omega})
reads
\be
\left( \frac{\partial \omega_0}{\partial m_b^*} \right )_{\mu_b, T}
= - \frac{\frac{g_{\omega b}} {m_\omega^2} \frac{\gamma}{(2\pi)^3} \frac 1T
\int d^3k  \frac{m_b^*}{\epsilon_b^*} [f_b(1-f_b)-\bar f_b (1- \bar f_b)]}
{1+ \frac{g_{\omega b}^2}{m_\omega^2}
\frac{\gamma}{(2\pi)^3} \frac 1T \int d^3k
[f_b(1-f_b) + \bar f_b (1-\bar f_b)]}.
\ee
Magnetic moments of baryons can be obtained by evaluating matrix
elements of the magnetic moment operator as
\begin{eqnarray}
\tilde{\bm \mu}_b &=& \left< \Psi_b \right|
~\sum_{i=q} \hat{\bm M}_i ~ \left| \Psi_b \right>,
\end{eqnarray}
where the sum is over the quarks in the bag, and $\left|\Psi_b
\right>$ is the wave function of a baryon $b$. 
The magnetic moment operator $\hat{\bm M}_i$ is given as
\begin{eqnarray}
\hat{\bm M}_i = \frac{\hat{Q}_i}{2}{\bm r}_i \times{\bm \alpha},
\label{eq:mag_op}
\end{eqnarray}
where $\hat{Q}_i$ and ${\bm r}_i$ are the charge and the position operators
of the $i$-th quark in the bag, and Dirac matrix ${\bm \alpha} = \gamma_0 {\bm \gamma}$.
Details for the evaluation of the matrix elements can be found
in Ref.~\cite{ryu09}, and we simply show the results for the analytic form
for the magnetic moment of each baryon :
\begin{eqnarray}
\mu_p(\rho, T) &=& \frac e2 D_u, ~~~~
\mu_n(\rho, T) = - \frac e3 D_u,  \nonumber \\
\mu_{\Lambda}(\rho, T) &=& - \frac e6 D_s, ~~~~
\mu_{\Sigma^+}(\rho, T) = \frac e6 \left[\frac 83 D_u +\frac 13 D_s\right], \nonumber \\
\mu_{\Sigma^0}(\rho, T) &=& \frac e6 \left[\frac 23 D_u +\frac 13 D_s\right], ~~~~
\mu_{\Sigma^-}(\rho, T) = \frac e6 \left[-\frac 43 D_u + \frac 13 D_s\right], \nonumber \\
\mu_{\Xi^0}(\rho, T) &=& -\frac e3 \left[\frac 13 D_u +\frac 23 D_s\right], ~~~~
\mu_{\Xi^-}(\rho, T) = \frac e6 \left[\frac 13 D_u -\frac 43 D_s\right].
\label{eq:mm}
\end{eqnarray}
where the integral $D_q$ is defined as
\begin{eqnarray}
D_q = \frac 43\, {\cal N}_q^2\, \beta_q
\left( \frac {R_b}{x_q} \right )^4 \int_0^{x_q}
y^3 j_0(y)j_1(y)dy.
\label{eq:Dq}
\end{eqnarray}
In the next section, we show numerical results and discuss them.

\section{Results and discussion}

\begin{figure}
\centering
\includegraphics[width=6.5cm]{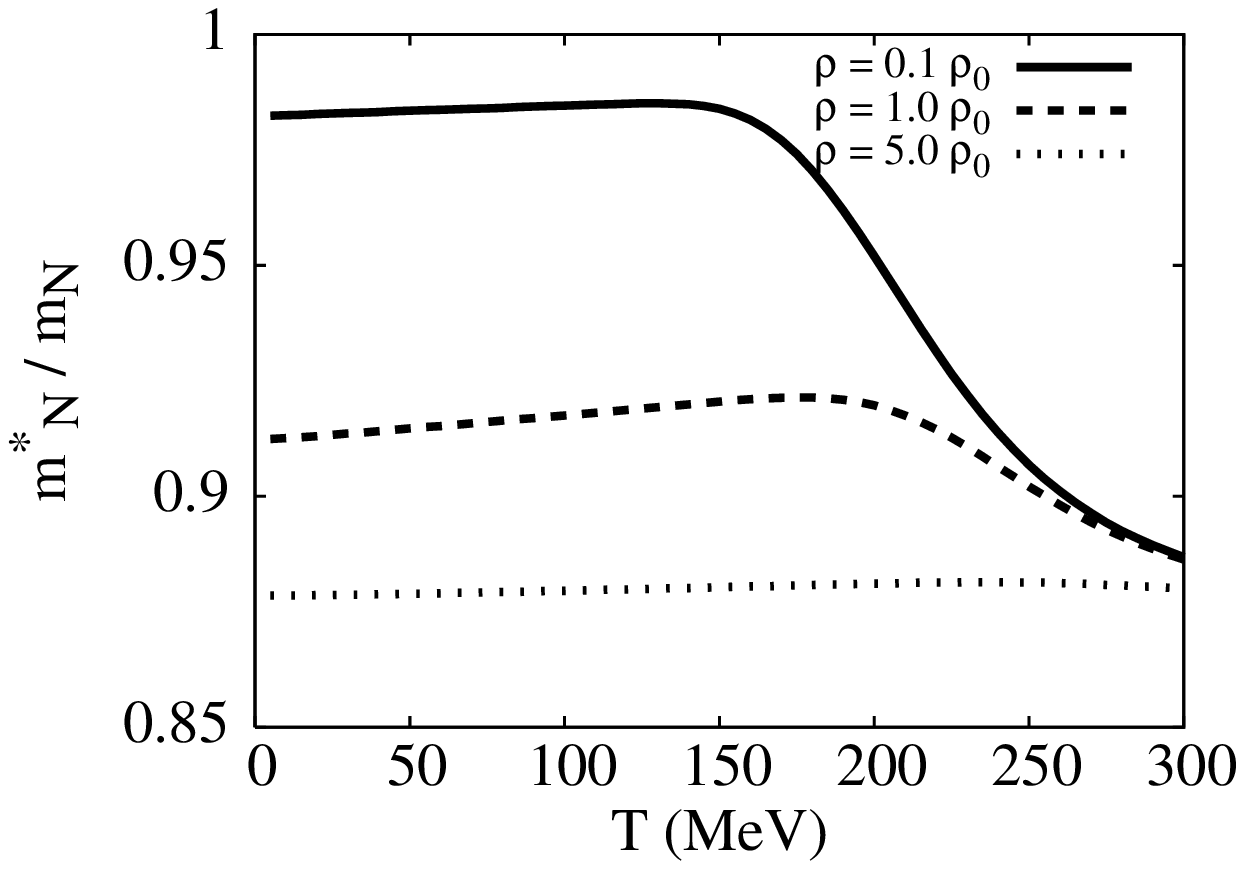}
\includegraphics[width=6.5cm]{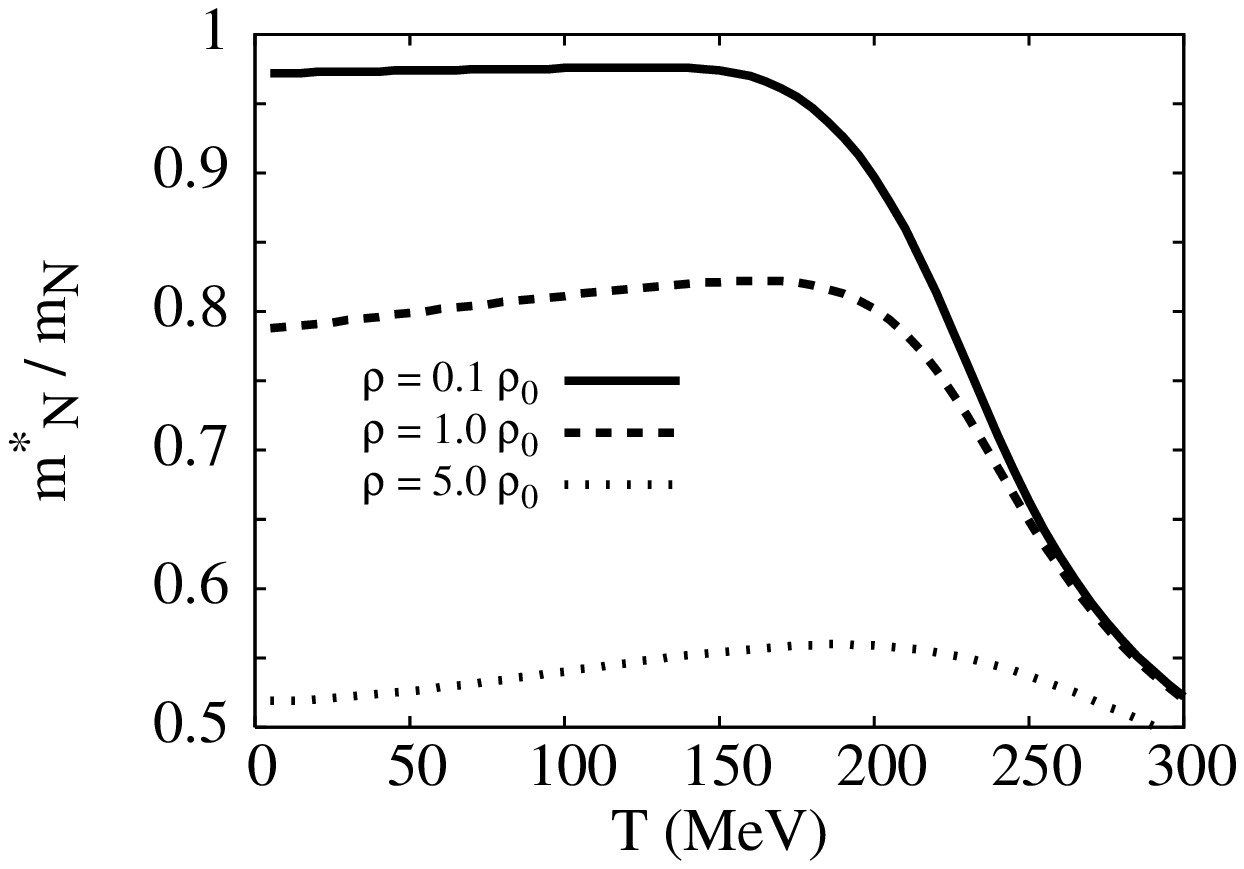}
\caption{Change of the nucleon mass with respect to
density and temperature in the QMC (left) and the MQMC (right) models.}
\label{fig:efmas}
\end{figure}

First, we show the nucleon mass as a function of density and
temperature in Fig.~\ref{fig:efmas}. At a temperature below
$100$ MeV, the nucleon mass decreases as density increases.
Exchange of the $\sigma$ meson mediates the scalar attraction. It
leads to the reduction of the nucleon mass at finite baryon
density. Since the effective mass of a baryon depends on $\sigma$
meson fields in Eq. (\ref{eq:efmass}), the effective mass of the
nucleon gets reduced. But the reduction is less in the QMC model than 
in the MQMC model, because the scalar-meson field in the QMC model is 
relatively smaller than the MQMC one.

The effect of temperature becomes clearer as density is lower. At
density $\rho/\rho_0 = 5$, temperature effect is almost invisible
in the QMC model, and it is a minor correction compared to the
density effect in the MQMC model. At $\rho/\rho_0 = 0.1$, change
of the mass is mainly driven by temperature, starting at
$T=150$~MeV in both QMC and MQMC models. The density and
temperature dependence of the nucleon mass in each model can be
understood by looking at the behavior of the $\sigma$-meson field
in Fig.2.

\begin{figure}
\centering
\includegraphics[width=6.5cm]{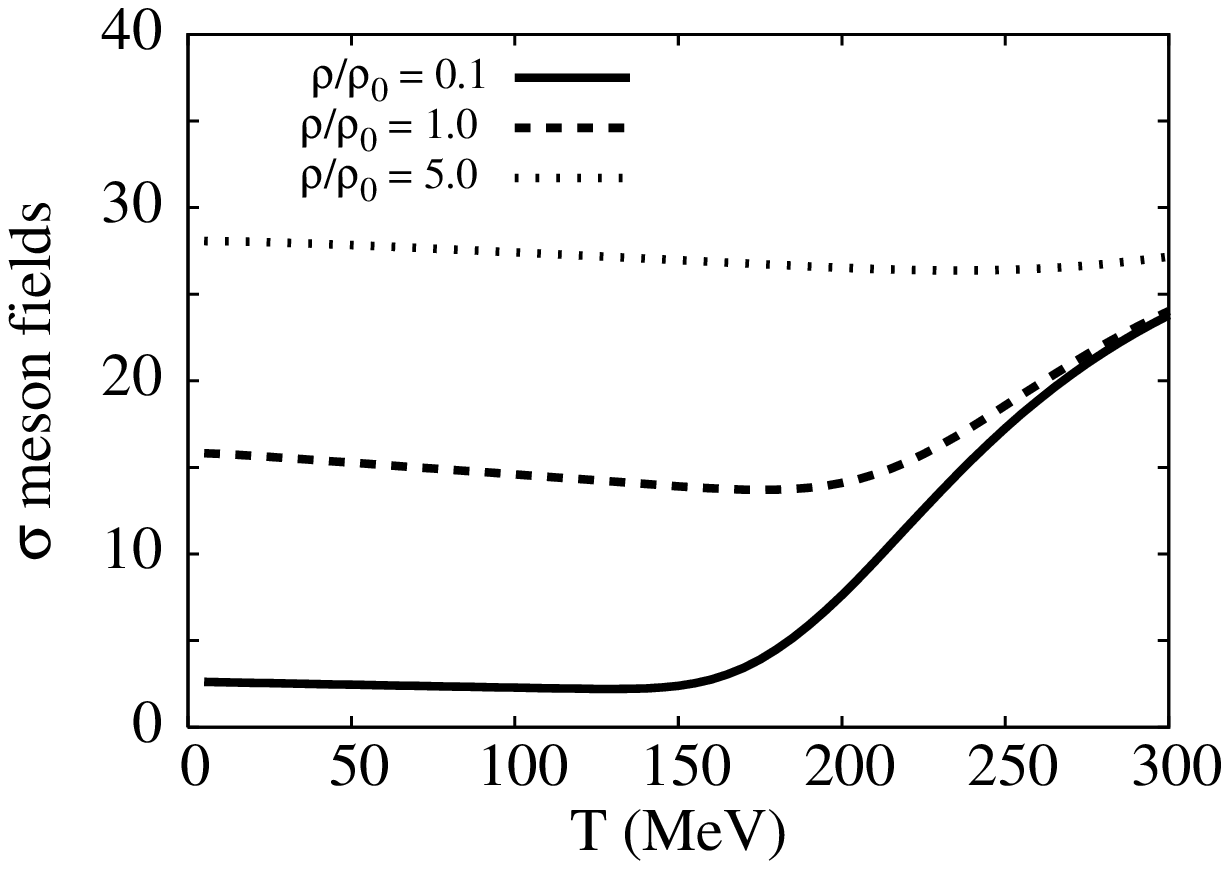}
\includegraphics[width=6.5cm]{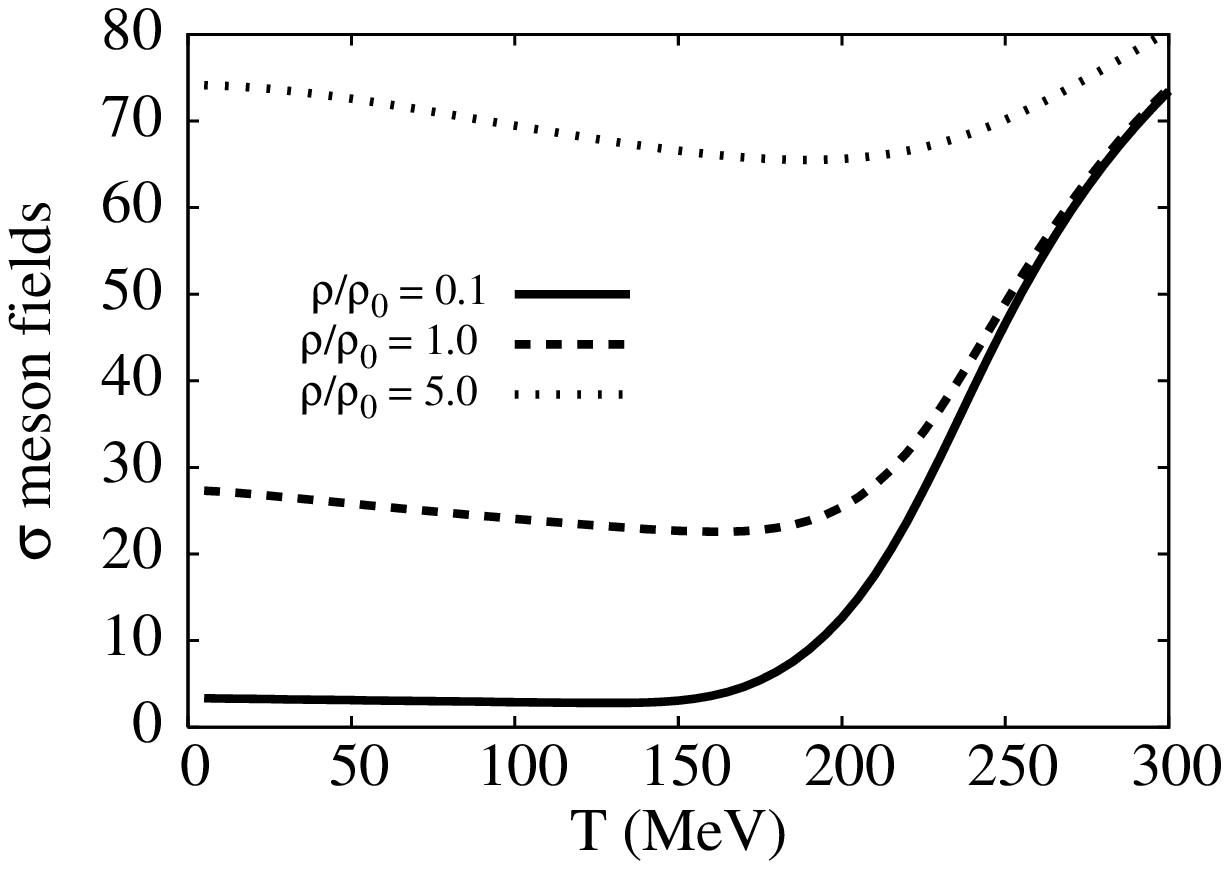}
\caption{Mean field of the $\sigma$ meson at various temperatures
and densities in the QMC(left) and the MQMC(right) models.}
\label{fig:sigma}
\end{figure}

Fig.~\ref{fig:sigma} shows the mean field values of $\sigma$-meson at
various densities and temperatures. One can note that shapes of
nucleon mass curves in Fig.~\ref{fig:efmas} are highly correlated
with the behavior of $\sigma$ field in both models. For instance, in
each model at $\rho/\rho_0 = 0.1$ and $1$, the nucleon mass converges
to a value as $T \to 300$~MeV. Similar convergence is observed
from the $\sigma$ field in both models. At small densities,
thermal excitation of nucleons and anti-nucleons is the main
source for the finite $\sigma$ value, and it consequently leads to
the decrease of the nucleon mass. Since the $\sigma$ field in the
MQMC model builds up more rapidly than in the QMC model, mass
reduction becomes drastic in the MQMC model.

\begin{figure}
\centering
\includegraphics[width=6.5cm]{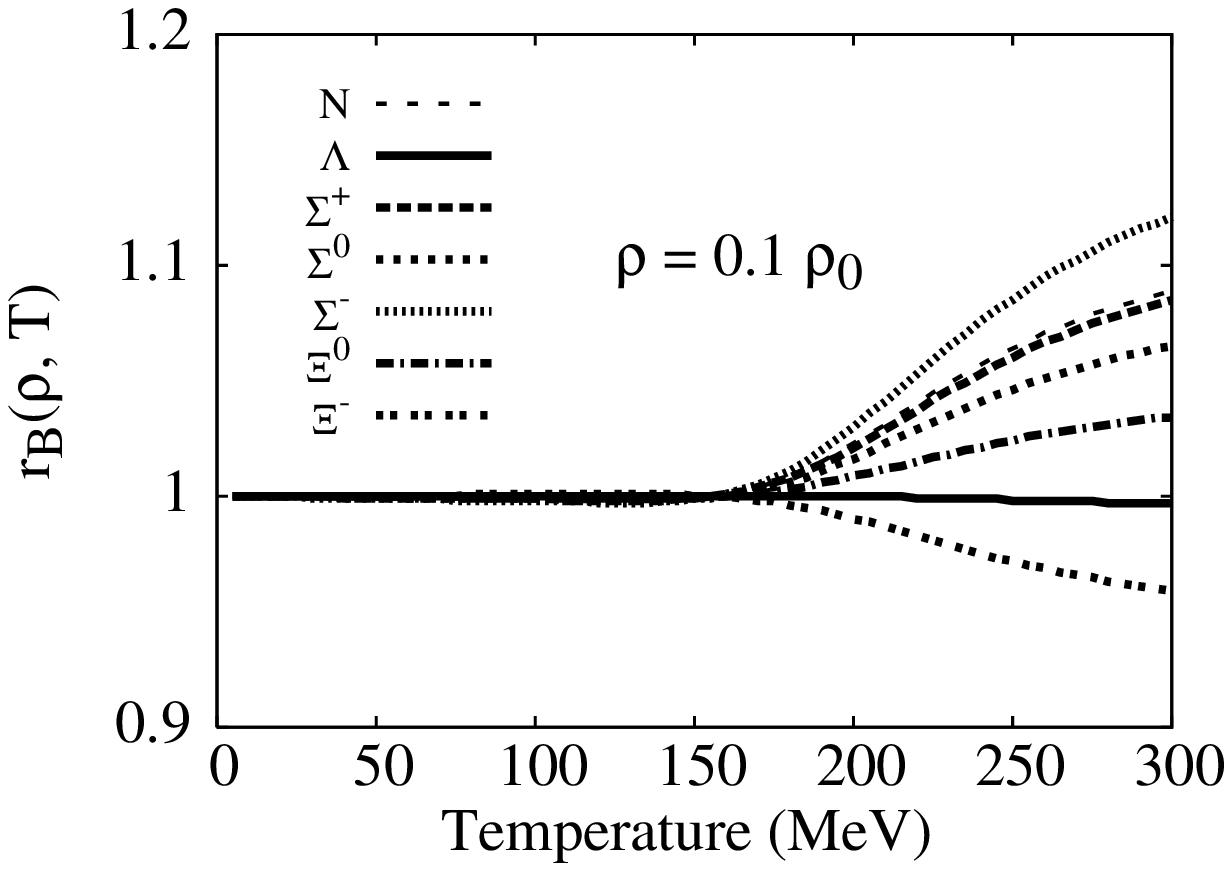}
\includegraphics[width=6.5cm]{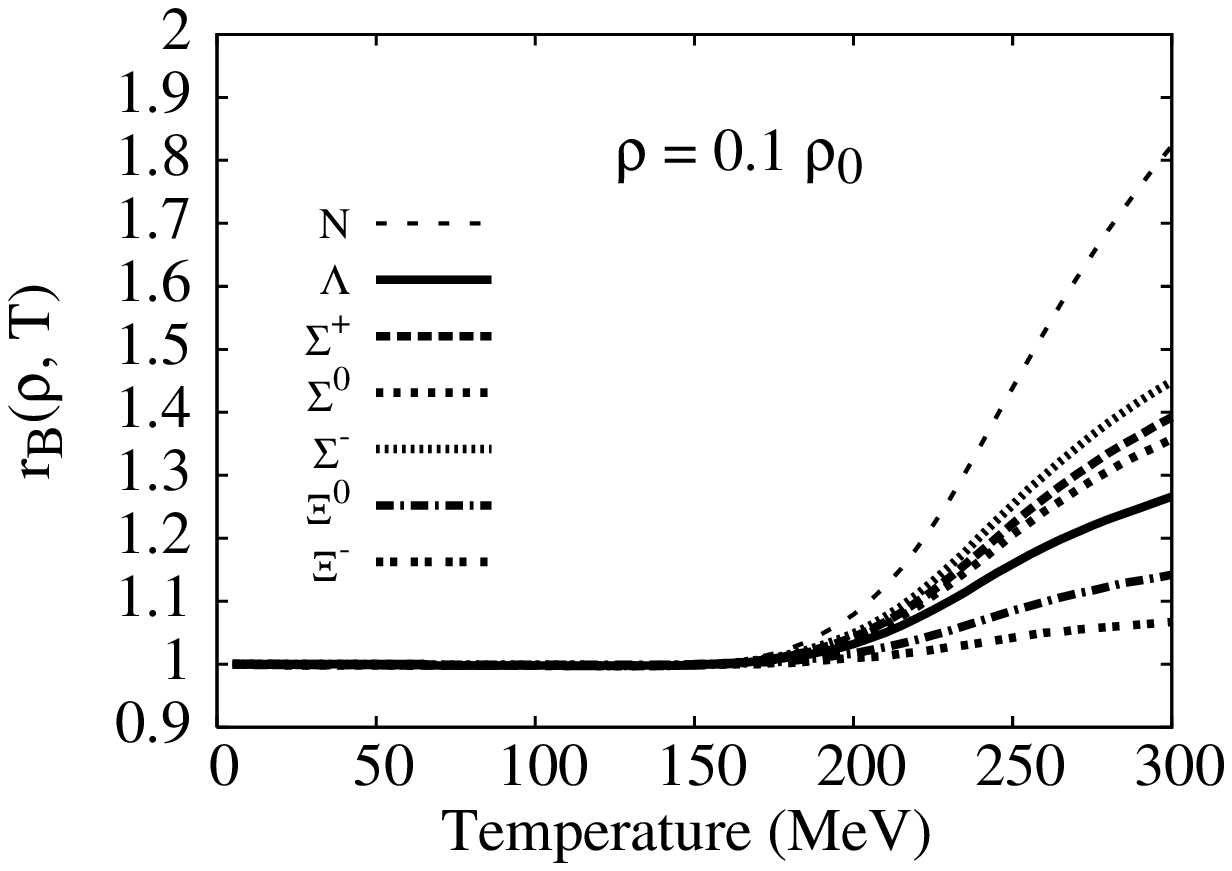} \\
\includegraphics[width=6.5cm]{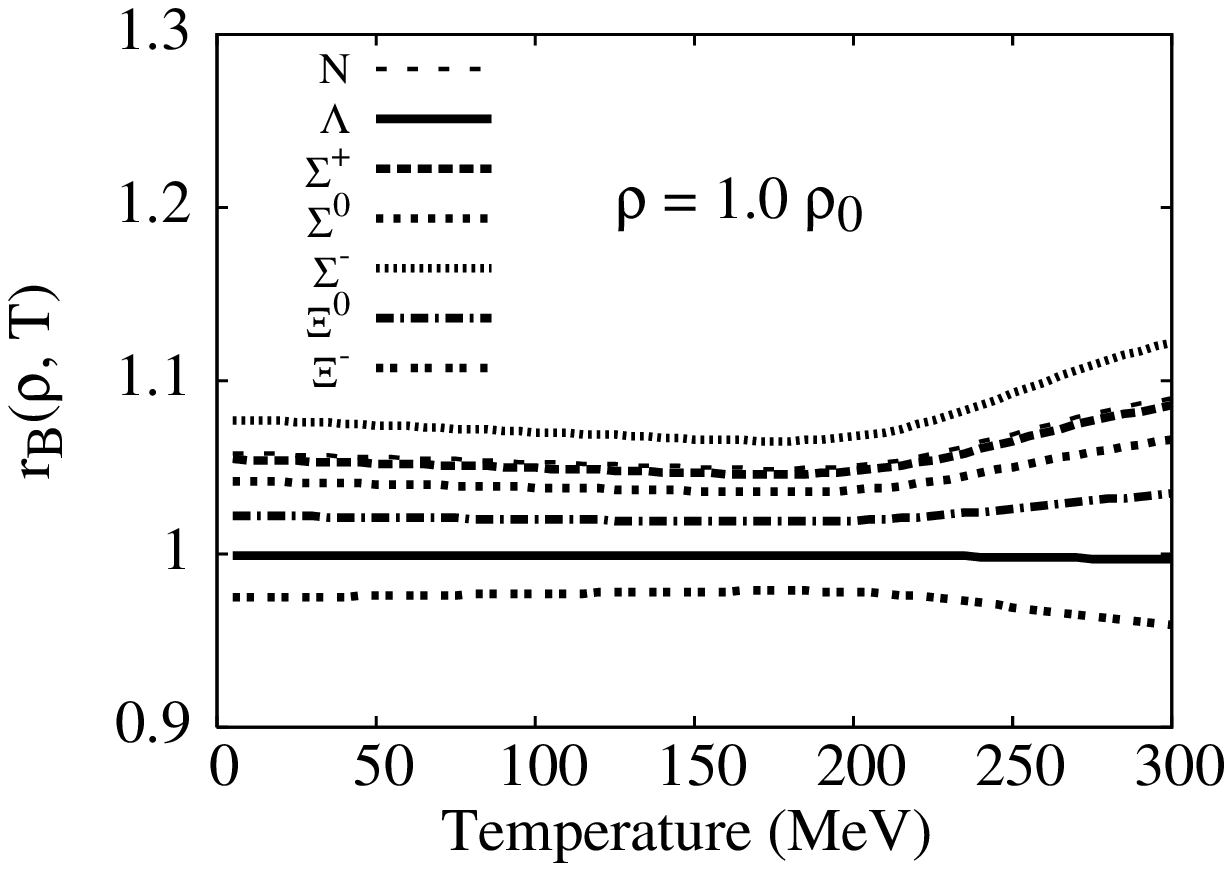}
\includegraphics[width=6.5cm]{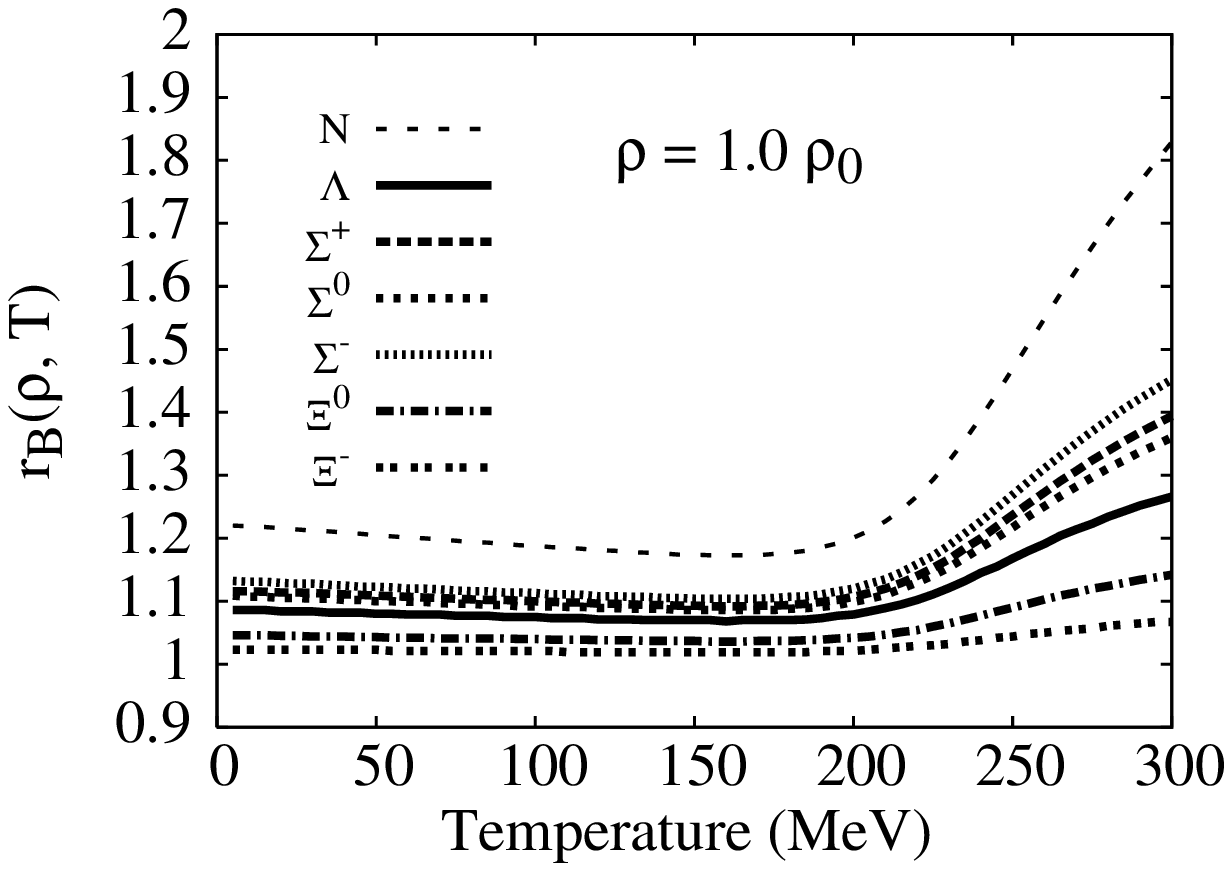} \\
\caption{Magnetic moments of octet baryons at $\rho/\rho_0 =
0.1$(up row) and $1$(down row) in the QMC(left column) and the
MQMC(right column) models. $r_b(\rho, T) \equiv {\mu_b(\rho, T)} /
{\mu_{b0}}$ where $\mu_{b0}$ is the magnetic moment of a baryon
$b$ at $\rho = 0$ and $T=0$. } 
\label{fig:mag}
\end{figure}

Fig.~\ref{fig:mag} compares magnetic moments of octet baryons in
the QMC and the MQMC model at $\rho/\rho_0=0.1$ and $1$, where
$r_b(\rho, T)$ is the ratio of the magnetic moment of a baryon $b$
in medium of density $\rho$ and temperature $T$ relative to its
free space value, \be r_b(\rho, T) \equiv \frac{\mu_b(\rho,
T)}{\mu_{b0}} \ee where $\mu_{b0}$ is the magnetic moment of a
baryon $b$ at $\rho = 0$ and $T=0$. At density close to zero,
temperature plays a dominant role in the change of the observable.
In both models, the temperature effect begins at around
$T=150$~MeV, where non-zero $\sigma$ field starts to have finite
values, and the change of the magnetic moment becomes sizable at
temperatures high enough.

\begin{figure}
\begin{center}
\epsfig{file=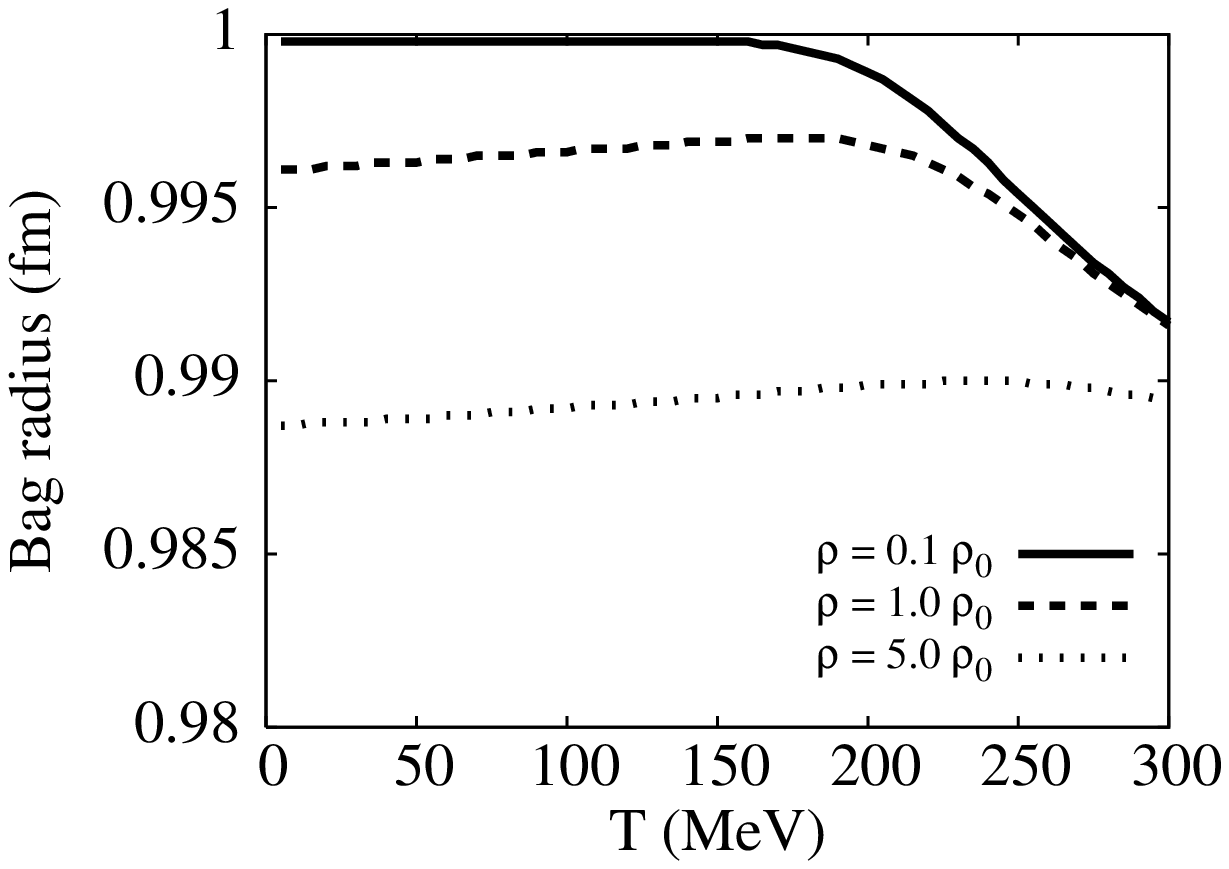,width=6.5cm}
\epsfig{file=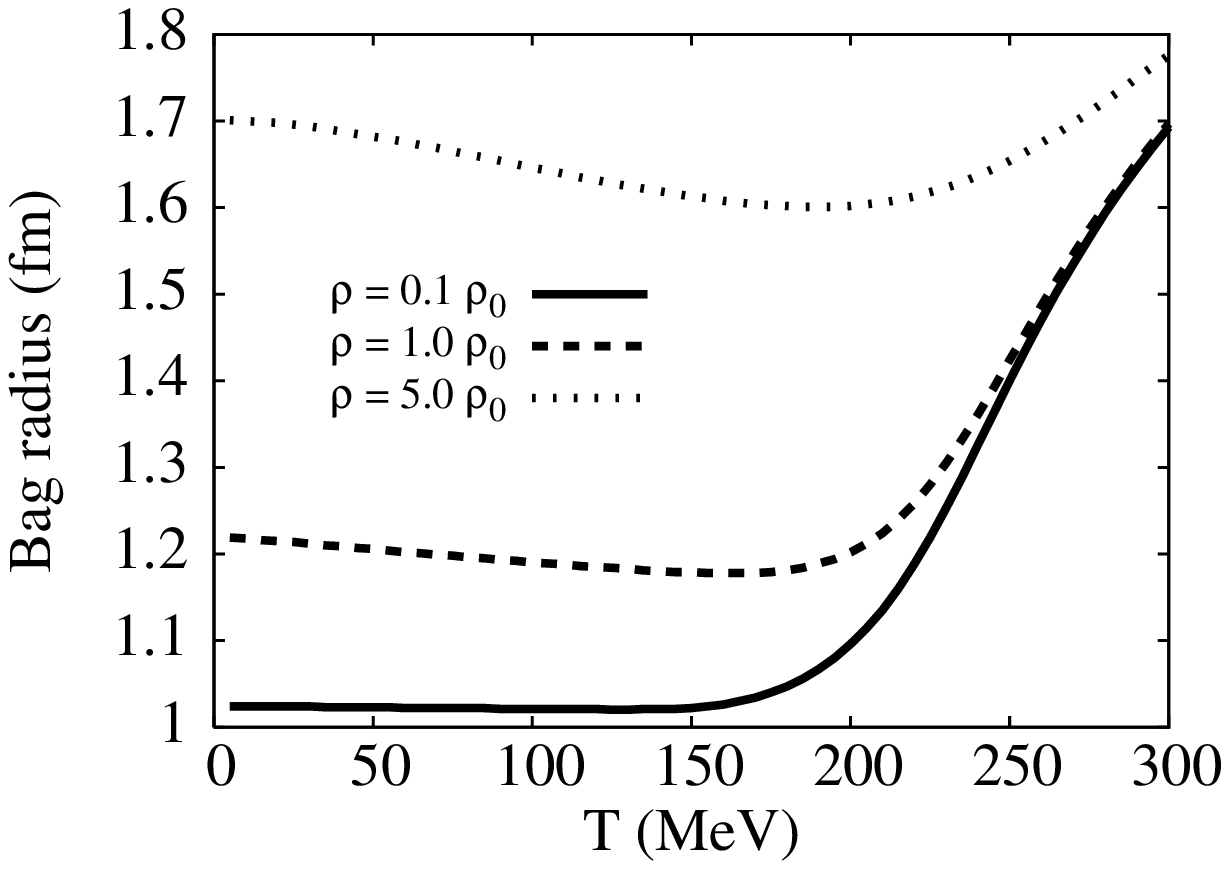,width=6.5cm}
\end{center}
\caption{Bag radius of the nucleon as a function of density
and temperature in the QMC model (left) and the MQMC one (right).}
\label{fig:bag_radius}
\end{figure}

Dependence on temperature is, however, contrastive in both models.
Finite $\sigma$ field causes the change of bag radius from its
free space value by the minimal condition of the mass,
Eq.~(\ref{eq:minimal}). In Ref.~\cite{ryu09}, we studied the
density effect to magnetic moments of octet baryons, and observed
a close correlation between the bag radius and magnetic moments.
Dependence on the temperature can be understood in a similar way
as shown in Fig.~\ref{fig:bag_radius}.

In Fig.~\ref{fig:bag_radius} we show the bag radius as a function
of temperature at $\rho/\rho_0 = 0.1,\, 1.0$ and $5.0$. At zero
temperature, the bag radius shrinks slighly from its free space
value as density increases in the QMC model. Temperature effect
shows a similar pattern, i.e. slightly decreases from free space
radius as the $\sigma$ field becomes finite at high temperatures.
In the MQMC model, on the other hand, overall shapes of the curves
for bag radius are very similar to those of the $\sigma$ field,
and they increase very quickly as temperature becomes high. As a
result, change of the magnetic moment at finite temperatures in
the MQMC model is more sensitive and significant than the QMC.

Another interesting result is the behavior of 
$r_{\Xi^-}$ in Fig.~\ref{fig:mag},
which decreases in the QMC model while it goes in the
opposite direction in the MQMC one.
Difference can be understood as follows.
In Eq.~(\ref{eq:mm}), one can see formulae for the magnetic momoments.
Since $\mu_p >0$, $\mu_\Lambda <0$ and $\mu_{\Xi^-}<0$ \cite{ryu09},
it is obvious that $D_{u0} >0$, $D_{s0} > 0$ and $D_{u0} - 4 D_{s0} <0$,
where 0 in the subscript denotes the value in free space.
In the QMC model, $r_N >1$ and $r_\Lambda <1$, which means that
$\Delta D_u >0$ and $\Delta D_s<0$.
Since $\Delta D_u - 4 \Delta D_s >0$ and $D_{u0} - 4 D_{s0}<0$,
we have $\Delta r_{\Xi^-}=(\Delta D_u - 4 \Delta D_s)/(D_{u0} - 4 D_{s0}) <0$ 
in the QMC model.
On the other hand, $\Delta D_u>0$ and $\Delta D_s>0$ in the MQMC model,
and thus the sign of $\Delta r_{\Xi^-}$ cannot be easily determined as
the QMC model.
The numerical result indicates that indeed $\Delta D_u - 4 \Delta D_s <0$,
and as a result $\Delta r_{\Xi^-}>0$ in the MQMC model.

Magnetic fields in the relativistic heavy-ion collision have been
recently considered, and relevant theories predict a field
strength as large as $1.5\times10^{19}$~G for the LHC energy
\cite{sokokov09}. Magnetic fields of order $10^{18}$~G are also
predicted in the core of the neutron star. With such high magnetic
fields, baryon masses and the equation of state become
substantially different from those without magnetic fields, and
they subsequently affects the critical density for the transition
to the deconfined quark matter and the maximum mass of the neutron
star \cite{wei06,rabhi09}.

Since a proton experiences the Landau quantization in the strong
magnetic field, energies of a neutron and a proton may have
different behaviors. However if we ignore the effect for a proton,
the nucleon energy at densities close to zero can be estimated as
\begin{eqnarray}
E_N \approx m_N - s \kappa_N B,
\end{eqnarray}
where $B$ is the magnetic field, $s$ is $+1$ ($-1$) for spin up
(down) and
$\kappa_N$ is defined as
$\kappa_b = (\mu_b / \mu_N - q_b m_p / m_b) \mu_N$ where $\mu_N$ is the magneton of a nucleon,
$\mu_N = e / (2m_p) = 3.15 \times 10^{-18}$ MeV G$^{-1}$.
In the free space, $m_N \simeq 940$~MeV
and $\kappa_p \simeq 1.79 \mu_N$ for a proton. Consequently, when
$B/B_c^e \sim 10^5$ G ($B_c^e = 4.414 \times 10^{13}$ G is the
critical electron field), the mass correction due to the magnetic
field, $|\kappa_p B | \simeq 25$ MeV, is about 3~\% of the free
nucleon mass.

At sufficiently high temperature, the situation may be
more drastic. 
We have shown in Figs.~1 and 3 that, as the temperature increases,
the effective mass of the octet baryons decreases, but their magnetic
moment increases from the free-space value.
Consequently, total corrections to the nucleon mass by the strong
magnetic fields at high temperature can be much larger than those
in free space.

For instance, in the MQMC model, at $\rho/\rho_0 = 0.1$ and
$T=250$~MeV, effective mass of the nucleon gets reduced as about
$0.65 m_N$. 
Net effects due to temperature and magnetic
fields give a correction of about 11~\% of the effective mass of
the nucleon, more than three times larger than that in free
space. If we only take the free space value for the magnetic
moment, then the correction due to the interaction with magnetic
fields is about 6~\%. Therefore, additional corrections
due to strong magnetic fields may exhibit clear discrimination of
finite temperature effects on the baryon mass and the magnetic
moment.

\section{Summary}
We have considered the change of magnetic moments of octet baryons
at a finite density and temperature. Two models, the quark-meson
coupling model and its modified one, are employed to investigate
any possible model dependence. Both models predict sizable effects
due to the variation of density and temperature.

Drastic changes of magnetic moments with the increase of
temperatures appear from about $T = 150 MeV$ at a low density, while
they change moderately at a high density. Consequently, if we
understand the sudden changes of magnetic moments as a signal of
the phase transition, our calculations may indicate the
first order phase transition around $T= 150$ MeV at low densities
and the second order phase transition at high densities. In
addition, at high temperature, the effective mass of a nucleon
for various densities is shown to converge to a value although
hadronic models used here should be scrutinized at such high
temperature.

As temperature increases, masses tend to decrease and magnetic
moments increase, although there exists non-negligible model
dependence on both models. If we consider strong magnetic fields
which can be realized in the relativistic heavy ion collision and
the neutron star, the change of magnetic moments and masses can
give more significant corrections to the nucleon energy. Such
changes of baryon properties at high temperature with the strong
magnetic field may give an insight into the phase transition to
the deconfinement and the restoration of broken symmetries. But,
since strong magnetic fields cause the Landau quantization of
charged particles and the breaking of spherical symmetry, more
detail calculations are to be done. Investigation along this
direction will be considered in near future.

\section*{ACKNOWLEDGMENTS}
Work of CHH was supported by a Korea Research Foundation Grant
funded by the Korean Government (MOEHRD, Basic Research Promotion
Fund) (2007-314-C00069).

\end{document}